# One-step implementation of a multi-target-qubit controlled-phase gate with photonic qubits encoded via eigenstates of the photon-number parity operator

Qi-Ping Su[1,*], Liang Bin,[1] Yu Zhang,[2,*] and Chui-Ping Yang[1,†]
[1]*Department of Physics, Hangzhou Normal University, Hangzhou 311121, China*
[2]*School of Physics, Nanjing University, Nanjing 210093, China*

In recent years, quantum state engineering and quantum information processing using microwave fields and photons have received increasing attention. In addition, multiqubit gates play an important role in quantum information processing. In this work, we propose to encode a photonic qubit via two arbitrary orthogonal eigenstates (with eigenvalues $\pm 1$, respectively) of the photon-number parity operator. With such encoding, we then present a single-step method to realize a multi-target-qubit controlled-phase gate with one photonic qubit simultaneously controlling $n-1$ target photonic qubits, by employing $n$ microwave cavities coupled to one superconducting flux qutrit. This proposal can be applied not only to implement *nonhybrid* multi-target-qubit controlled-phase gates using photonic qubits with various encodings, but also to realize *hybrid* multi-target-qubit controlled-phase gates using photonic qubits with different encodings. The gate realization requires only a single-step operation. The gate operation time does not increase with the number of target qubits. Because the qutrit remains in the ground state during the entire operation, decoherence from the qutrit is greatly suppressed. As an application, we show how to apply this gate to generate a multicavity Greenberger-Horne-Zeilinger (GHZ) entangled state with general expression. Depending on the specific encodings, we further discuss the preparation of several nonhybrid and hybrid GHZ entangled states of multiple cavities. We numerically investigate the circuit-QED experimental feasibility of creating a three-cavity spin-coherent hybrid GHZ state. This proposal can be extended to accomplish the same tasks in a wide range of physical systems, such as multiple microwave or optical cavities coupled to a three-level natural or artificial atom.

## I. INTRODUCTION AND MOTIVATION

Quantum computing has attracted much interest as quantum computers can in principle solve hard computing problems much more efficiently than classical computers. Multiqubit gates are particularly attractive and have been considered as appealing building blocks for quantum computing. As is well known, there exist two kinds of important multiqubit gates, namely, *multi-control-qubit* gates with multiple control qubits acting on a single target qubit, and *multi-target-qubit* gates with a single qubit simultaneously controlling multiple target qubits. These two kinds of multiqubit gates have many applications in quantum information processing (QIP). For instance, they are key elements in quantum algorithms [1–3], quantum Fourier transform [4], quantum cloning [5], error correction [6–8], and entanglement preparation [9]. Therefore, it is necessary and important to implement these two kinds of multiqubit gates.

A multiqubit gate can in principle be constructed by using single-qubit and two-qubit fundamental gates. However, when the conventional gate-decomposition protocols are used to construct a multiqubit gate [10–13], the number of fundamental gates increases substantially and the procedure usually becomes complicated with increasing the number of qubits. As a result, the operation time required to implement a multiqubit gate will be quite long and thus the fidelity will be significantly deteriorated by decoherence. Therefore, it is worth seeking efficient ways to *directly* implement multiqubit gates. In the past years, a number of proposals have been put forward for directly realizing multi-control-qubit gates and multi-target-qubit gates using *matter* qubits (e.g., atoms, trapped ions, superconducting qubits, quantum dots, nitrogen-vacancy centers, etc.) [14–33].

Circuit quantum electrodynamics (QED), composed of microwave cavities and superconducting (SC) qubits, has been considered as one of the best platforms for quantum computing [34–43]. SC qubits (such as flux qubits, transmon qubits, Xmon qubits, and fluxonium qubits) and microwave resonators (a.k.a. cavities) can be fabricated using modern integrated circuit technology. Because their energy-level spacings can be rapidly adjusted *in situ* [44–46] and their coherence time has been much improved [47–51], SC qubits play an important role in QIP. In addition, due to the high-quality factors of microwave cavities demonstrated in experiments [52–56], the microwave photons contained in microwave cavities or resonators can have a long lifetime [57]. Recently, quantum state engineering and QIP using microwave fields and photons have received increasing attention. In this fast-growing field, photonic qubits, which are encoded via discrete-variable or continuous-variable states of microwave cavity fields, have been adopted as information carriers for quantum computing and communication.

---
[*]These authors contributed equally to this work.
[†]yangcp@hznu.edu.cn

Compared to SC qubits or other matter qubits, photonic qubits can have various encodings. Each encoding pattern has its own advantage and can be applied in different tasks of QIP. For example, photonic qubits, encoded via coherent states, are tolerant to single-photon loss [58]; the lifetime of photonic qubits encoded via cat states can be greatly enhanced by quantum error correction [59]; and photonic qubits encoded via the vacuum state and the single photon state are relatively easy to manipulate [60]. Over the past years, a number of proposals have been presented for directly realizing single-qubit gates, two-qubit gates, multi-control-qubit gates, and multi-target-qubit gates, by using photonic qubits which are specifically encoded via (i) coherent states [61,62], (ii) cat states [63–68], (iii) the vacuum and single-photon states [69–73], or (iii) polarization, spatial, and temporal degrees of freedom (DOFs) of photons [74–80], etc. The previous works have made great contributions to quantum computing and QIP with photonic qubits. Generally speaking, different encodings of photonic qubits may require different methods in implementing quantum gates with photonic qubits. In other words, methods for realizing quantum gates using photonic qubits with specific encodings may not be applied to implement quantum gates using photonic qubits with various encodings.

Motivated by the above, in this work, our goal is to propose an idea to encode photonic qubits, i.e., encoding a photonic qubit via two *arbitrary* orthogonal eigenstates $|\varphi_e\rangle$ and $|\varphi_o\rangle$ of the photon-number parity operator $\hat{\pi} = e^{i\pi \hat{a}^+ \hat{a}}$. Here, for the state $|\varphi_e\rangle$, the photon-number parity operator $\hat{\pi}$ has an eigenvalue 1, while for the state $|\varphi_o\rangle$, the photon-number parity operator $\hat{\pi}$ has an eigenvalue $-1$. After a deep search of literature, we find no indication that this idea for encoding the photonic qubit has been reported yet.

With such encoding, we then propose a one-step method to implement a multi-target-qubit controlled-phase gate with one photonic qubit simultaneously controlling $(n-1)$ target photonic qubits, based on circuit QED. This gate is realized by employing $n$ cavities coupled to a superconducting qutrit. As discussed in Sec. III C below, this proposal can be applied to photonic qubits with various encodings. More remarkably, as discussed in Sec. III D below, the proposal can be used to realize multi-target-qubit *hybrid* controlled-phase gates using photonic qubits with different encodings. Hybrid quantum gates are of fundamental interest in quantum physics and have significant applications in hybrid quantum communication and quantum computation.

Greenberger-Horne-Zeilinger (GHZ) entangled states are of great interest in the foundations of quantum mechanics [81], and have many important applications in QIP [82], quantum communications [83], quantum metrology [84], error-correction protocols [85], and high-precision spectroscopy [86]. On the other hand, hybrid entangled states play a key role in QIP and quantum technology. They can serve as important quantum channels and intermediate resources for quantum technologies, covering the transmission, operation, and storage of quantum information between different formats and encodings [87–89]. In this work, as an application of the proposed gate, we further discuss how to generate a multicavity GHZ entangled state with general expression. Depending on the specific encodings $|\varphi_e\rangle$ and $|\varphi_o\rangle$, we also discuss the preparation of several nonhybrid and hybrid GHZ entangled states of multiple cavities. Furthermore, we numerically investigate the circuit-QED experimental feasibility of preparing a spin-coherent hybrid GHZ state of three cavities.

We believe that this work is interesting and important from these aspects: First, the proposal can be applied not only to implement nonhybrid multi-target-qubit controlled-phase gates using photonic qubits with various encodings, but also to realize hybrid multi-target-qubit controlled-phase gates using photonic qubits with different encodings. Second, the gate implementation is very simple because it requires only a single-step operation that is independent of the number of target qubits, while the gate realization becomes complicated (especially for a large number of target qubits) by the use of the conventional gate-decomposition protocols [10–13]. Third, the gate operation time does not depend on the number of target qubits and thus does not increase with the number of target qubits; however, the operation time required to realize this multi-target-qubit gate will increase with the number of target qubits when the conventional gate-decomposition protocols [10–13] are applied. Last, the qutrit stays in the ground state during the entire gate operation, thus decoherence from the qutrit is significantly suppressed.

This paper is organized as follows. In Sec. II, we give a brief introduction to the multi-target-qubit gate considered in this work. In Sec. III, we explicitly show how to realize this multi-target-qubit gate, introduce concrete examples for the photonic-qubit encodings, and give a brief discussion on the nonhybrid and hybrid multi-target-qubit gates. In Sec. IV, we show how to apply this gate to generate a multicavity GHZ entangled state with general expression. Depending on the specific encodings $|\varphi_e\rangle$ and $|\varphi_o\rangle$, we further show the generation of several nonhybrid and hybrid GHZ entangled states of multiple cavities. In Sec. V, we numerically analyze the circuit-QED experimental feasibility of creating a three-cavity spin-coherent hybrid GHZ state. A concluding summary is given in Sec. VI.

## II. THE MULTI-TARGET-QUBIT CONTROLLED-PHASE GATE

For $n$ qubits, there exist $2^n$ computational basis states, which form a set of complete orthogonal bases in a $2^n$-dimensional Hilbert space of the $n$ qubits. An $n$-qubit computational basis state is denoted as $|i_1\rangle|i_2\rangle|i_3\rangle \cdots |i_n\rangle$, where subscripts 1, 2, ..., $n$ represent qubits 1, 2, ..., $n$, respectively, and $i_j \in \{0, 1\}$ ($j = 1, 2, ..., n$). The multi-target-qubit gate considered in this work (Fig. 1) is described by

$$\begin{aligned}
|0_1\rangle|i_2\rangle|i_3\rangle \cdots |i_n\rangle &\rightarrow |0_1\rangle|i_2\rangle|i_3\rangle \cdots |i_n\rangle, \\
|1_1\rangle|i_2\rangle|i_3\rangle \cdots |i_n\rangle &\rightarrow |1_1\rangle(-1)^{i_2}(-1)^{i_3} \cdots \\
&\quad (-1)^{i_n}|i_2\rangle|i_3\rangle \cdots |i_n\rangle,
\end{aligned} \quad (1)$$

which shows that if the control qubit (qubit 1) is in state $|0\rangle$, the states of each of the $(n-1)$ target qubits (qubits 2, 3, ..., $n$) remain unchanged, while if the control qubit is in state $|1\rangle$, state $|1\rangle$ of each target qubit undergoes a phase flip from sign $+$ to $-$.

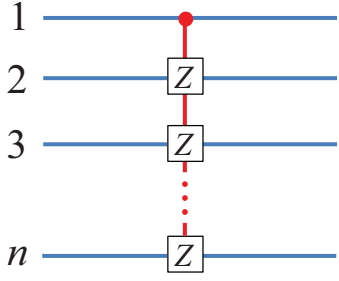

FIG. 1. Circuit of a multi-target-qubit controlled phase gate with one control qubit (qubit 1) simultaneously controlling $n-1$ target qubits $(2, 3, ..., n)$. When the control qubit (on the filled circle) is in state $|1\rangle$, state $|1\rangle$ at $Z$ for each target qubit is phase flipped as $|1\rangle \rightarrow -|1\rangle$ but nothing happens to state $|0\rangle$ at $Z$ for each target qubit. In this work, each qubit is a photonic qubit, for which the two logic states are encoded through two *arbitrary* orthogonal eigenstates (with eigenvalues $\pm 1$, respectively) of the photon-number parity operator.

In this work, the $n$ qubits involved in this multiqubit gate are photonic qubits. For photonic qubit $k$ ($k = 1, 2, ..., n$), the two logical states $|0_k\rangle$ and $|1_k\rangle$ are encoded with two arbitrary orthogonal eigenstates $|\varphi_e\rangle_{c_k}$ (with eigenvalue 1) and $|\varphi_o\rangle_k$ (with eigenvalue $-1$) of the photon-number parity operator $\hat{\pi}_k = e^{i\pi \hat{a}_k^+ \hat{a}_k}$ of cavity $k$ ($k = 1, 2, ..., n$), i.e.,

$$|0_k\rangle = |\varphi_e\rangle_{c_k} = \sum_{p_k} d_{p_k} |p_k\rangle,$$

$$|1_k\rangle = |\varphi_o\rangle_{c_k} = \sum_{q_k} d_{q_k} |q_k\rangle, \quad (2)$$

where the subscript $c_k$ represents cavity $k$ ($k = 1, 2, ..., n$), $|p_k\rangle$ is a Fock state of cavity $k$ with any even number $p_k$ of photons, while $|q_k\rangle$ is a Fock state of cavity $k$ with any odd number $q_k$ of photons. The sum $\sum_{p_k}$ is taken over all Fock states with even-number photons, while the sum $\sum_{q_k}$ is taken over all Fock states with odd-number photons. The coefficients $d_{p_k}$ and $d_{q_k}$ satisfy the normalization conditions $\sum_{p_k} |d_{p_k}|^2 = \sum_{q_k} |d_{q_k}|^2 = 1$. Obviously, the two states $|\varphi_e\rangle_{c_k}$ and $|\varphi_o\rangle_{c_k}$ are orthogonal to each other. One can easily check $\hat{\pi}_k |\varphi_e\rangle_{c_k} = |\varphi_e\rangle_{c_k}$ and $\hat{\pi}_k |\varphi_o\rangle_{c_k} = -|\varphi_o\rangle_{c_k}$, namely, the two states $|\varphi_e\rangle_{c_k}$ and $|\varphi_o\rangle_{c_k}$ are the eigenstates of the photon-number parity operator $\hat{\pi}_k$ with eigenvalues 1 and $-1$, respectively. Please keep in mind that, in Eq. (2), $p_k$ is a nonnegative even number while $q_k$ is a positive odd number, which will apply in the next section.

## III. IMPLEMENTATION OF THE MULTI-TARGET-QUBIT CONTROLLED-PHASE GATE

In this section, we will show how to implement the multi-target-qubit controlled-phase gate described by Eq. (1). We then discuss some issues related to the gate implementation, provide examples for the photonic qubit encodings, and give a brief discussion on the nonhybrid and hybrid multi-target-qubit gates.

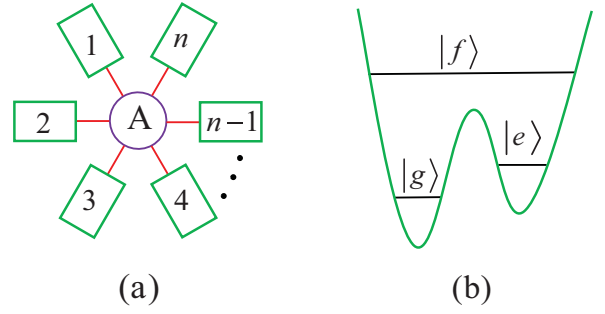

FIG. 2. (a) Schematic circuit of $n$ microwave cavities coupled to a SC flux qutrit. Each square represents a one-dimensional (1D) or three-dimensional (3D) microwave cavity. The circle $A$ represents the flux qutrit, which is inductively or capacitively coupled to each cavity. (b) Level configuration of the flux qutrit, for which the transition between the two lowest levels can be made weak by increasing the barrier between two potential wells.

### A. The gate realization

Consider a physical system consisting of $n$ microwave cavities $(1, 2, ..., n)$ coupled to a SC flux qutrit [Fig. 2(a)]. The three levels of the qutrit are labeled as $|g\rangle$, $|e\rangle$, and $|f\rangle$ [Fig. 2(b)]. The $|g\rangle \leftrightarrow |e\rangle$ transition can be made weak by increasing the barrier between the two potential wells. Suppose that cavity 1 is dispersively coupled to the $|g\rangle \leftrightarrow |f\rangle$ transition with coupling constant $g_1$ and detuning $\delta_1$ but highly detuned (decoupled) from the $|e\rangle \leftrightarrow |f\rangle$ transition of the qutrit. In addition, assume that cavity $l$ ($l = 2, 3, ..., n$) is dispersively coupled to the $|e\rangle \leftrightarrow |f\rangle$ transition with coupling constant $g_l$ and detuning $\delta_l$ but highly detuned (decoupled) from the $|g\rangle \leftrightarrow |f\rangle$ transition of the qutrit (Fig. 3). These conditions can be met by prior adjustment of the level spacings of the qutrit or the frequencies of the cavities. Note that both of the level spacings of a SC qutrit and the frequency of a microwave cavity can be rapidly tuned within a few nanoseconds [44–46,90,91].

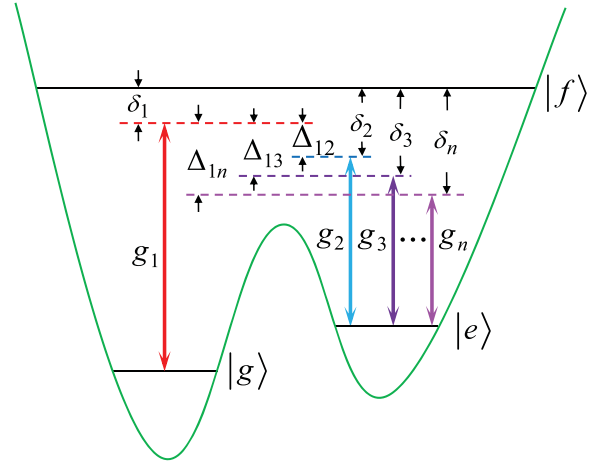

FIG. 3. Cavity 1 is dispersively coupled to the $|g\rangle \leftrightarrow |f\rangle$ transition of the qutrit with coupling strength $g_1$ and detuning $\delta_1$, while cavity $l$ ($l = 2, 3, ..., n$) is dispersively coupled to the $|e\rangle \leftrightarrow |f\rangle$ transition of the qutrit with coupling strength $g_l$ and detuning $\delta_l$. Here, $\Delta_{1l} = \delta_l - \delta_1$ ($l = 2, 3, ..., n$).

After the above considerations, the Hamiltonian of the whole system in the interaction picture and under the rotating-wave approximation is given by (in units of $\hbar = 1$)

$$H_I = g_1(e^{-i\delta_1 t}\hat{a}_1^+ \sigma_{fg}^- + \text{H.c.}) + \sum_{l=2}^{n} g_l(e^{-i\delta_l t}\hat{a}_l^+ \sigma_{fe}^- + \text{H.c.}), \quad (3)$$

where $\hat{a}_1$ ($\hat{a}_l$) is the photon annihilation operator of cavity 1 ($l$), $\sigma_{fg}^- = |g\rangle\langle f|$, $\sigma_{fe}^- = |e\rangle\langle f|$, $\delta_1 = \omega_{fg} - \omega_{c_1} > 0$, and $\delta_l = \omega_{fe} - \omega_{c_l} > 0$ (Fig. 3). Here, $\omega_{fg}$ ($\omega_{fe}$) is the $|f\rangle \leftrightarrow |g\rangle$ ($|f\rangle \leftrightarrow |e\rangle$) transition frequency of the qutrit, while $\omega_{c_1}$ ($\omega_{c_l}$) is the frequency of cavity 1 ($l$).

Under the large detuning conditions $\delta_1 \gg g_1$ and $\delta_l \gg g_l$ ($l = 2, 3, ..., n$), the energy exchange between the coupler qutrit and the cavities is negligible. In addition, under the condition of

$$\frac{|\delta_p - \delta_q|}{\delta_p^{-1} + \delta_q^{-1}} \gg g_p g_q \quad (4)$$

(where $p, q \in \{2, 3, ..., n\}$ and $p \neq q$), the interaction between the cavities (2, 3, ..., $n$), induced by the coupler qutrit, is negligible. Therefore, the Hamiltonian (3) becomes [92–94]

$$H_e = -\lambda_1 \hat{n}_1 |g\rangle\langle g| + \lambda_1(1+\hat{n}_1)|f\rangle\langle f|$$
$$- \sum_{l=2}^{n} \lambda_l \hat{n}_l |e\rangle\langle e| + \sum_{l=2}^{n} \lambda_l(1+\hat{n}_l)|f\rangle\langle f|$$
$$- \sum_{l=2}^{n} \lambda_{1l}(e^{i\Delta_{1l}t}\hat{a}_1^+ \hat{a}_l \sigma_{eg}^- + \text{H.c.}), \quad (5)$$

where $\sigma_{eg}^- = |g\rangle\langle e|$, $\lambda_1 = g_1^2/\delta_1$, $\lambda_l = g_l^2/\delta_l$, $\lambda_{1l} = (g_1 g_l/2)(1/\delta_1 + 1/\delta_l)$, and $\Delta_{1l} = \delta_l - \delta_1 = \omega_{c_1} - \omega_{c_l} - \omega_{eg} > 0$. In Eq. (5), the terms in the first two lines describe the photon-number-dependent stark shifts of the energy levels $|g\rangle$, $|e\rangle$, and $|f\rangle$, while the terms in the last line describe the $|g\rangle \leftrightarrow |e\rangle$ coupling caused due to the cooperation of cavities 1 and $l$ ($l = 2, 3, ..., n$). For $\Delta_{1l} \gg \{\lambda_1, \lambda_l, \lambda_{1l}\}$, the effective Hamiltonian $H_e$ turns into [92–94]

$$H_e = -\lambda_1 \hat{n}_1 |g\rangle\langle g| + \lambda_1(1+\hat{n}_1)|f\rangle\langle f|$$
$$- \sum_{l=2}^{n} \lambda_l \hat{n}_l |e\rangle\langle e| + \sum_{l=2}^{n} \lambda_l(1+\hat{n}_l)|f\rangle\langle f|$$
$$+ \sum_{l=2}^{n} \chi_{1l}\hat{n}_1(1+\hat{n}_l)|g\rangle\langle g| - \sum_{l=2}^{n} \chi_{1l}(1+\hat{n}_1)\hat{n}_l|e\rangle\langle e|, \quad (6)$$

where $\chi_{1l} = \lambda_{1l}^2/\Delta_{1l}$. In the case of the levels $|e\rangle$ and $|f\rangle$ being initially not occupied, these levels will not be populated because neither $|g\rangle \to |e\rangle$ transition nor $|g\rangle \to |f\rangle$ transition is induced by the Hamiltonian (6). Therefore, the Hamiltonian (6) reduces to

$$H_e = -\lambda_1 \hat{n}_1 |g\rangle\langle g| + \sum_{l=2}^{n} \chi_{1l}\hat{n}_1(1+\hat{n}_l)|g\rangle\langle g|. \quad (7)$$

In the following, we assume that the qutrit is initially in the ground state $|g\rangle$. Note that the qutrit will remain in this state because none of interlevel transitions of the qutrit can be induced by the Hamiltonian (7). The Hamiltonian (7) thus reduces to

$$\widetilde{H}_e = \eta \hat{n}_1 + \sum_{l=2}^{n} \chi_{1l}\hat{n}_1 \hat{n}_l, \quad (8)$$

where $\eta = -\lambda_1 + \sum_{l=2}^{n} \chi_{1l}$. The effective Hamiltonian $\widetilde{H}_e$ characterizes the dynamics of the cavity system. Under this Hamiltonian, the unitary operator $U = e^{-i\widetilde{H}_e t}$ describing the state time evolution of the cavity system can be expressed as

$$U = U_1 \otimes \prod_{l=2}^{n} U_{1l}, \quad (9)$$

with

$$U_1 = \exp(-i\eta \hat{n}_1 t), \quad (10)$$

$$U_{1l} = \exp(-i\chi_{1l}\hat{n}_1 \hat{n}_l t), \quad (11)$$

where $U_1$ is a unitary operator on cavity 1 while $U_{1l}$ is a unitary operator on cavities 1 and $l$ ($l = 2, 3, ..., n$).

As stated in the previous section [see Eq. (2) and the discussion there], for photonic qubit $k$ ($k = 1, 2, ..., n$), the two logical states $|0_k\rangle$ and $|1_k\rangle$ are encoded with two arbitrary orthogonal eigenstates $|\varphi_e\rangle_{c_k}$ and $|\varphi_o\rangle_k$ of the photon-number parity operator $\hat{\pi}_k = e^{i\pi \hat{a}_k^+ \hat{a}_k}$ of cavity $k$. For two photonic qubits 1 and $l$ ($l = 2, 3, ..., n$), there are four computational basis states, denoted by $|0_1 0_l\rangle$, $|0_1 1_l\rangle$, $|1_1 0_l\rangle$, and $|1_1 1_l\rangle$. According to Eqs. (11) and (2), it is easy to verify that the unitary operator $U_{1l}$ acting on the four basis states results in the following state transformation:

$$U_{1l}|0_1 0_l\rangle = \sum_{p_1, p_l} \exp(-ip_1 p_l \chi_{1l} t) d_{p_1} d_{p_l} |p_1\rangle|p_l\rangle,$$
$$U_{1l}|0_1 1_l\rangle = \sum_{p_1, q_l} \exp(-ip_1 q_l \chi_{1l} t) d_{p_1} d_{q_l} |p_1\rangle|q_l\rangle,$$
$$U_{1l}|1_1 0_l\rangle = \sum_{q_1, p_l} \exp(-iq_1 p_l \chi_{1l} t) d_{q_1} d_{p_l} |q_1\rangle|p_l\rangle,$$
$$U_{1l}|1_1 1_l\rangle = \sum_{q_1, q_l} \exp(-iq_1 q_l \chi_{1l} t) d_{q_1} d_{q_l} |q_1\rangle|q_l\rangle. \quad (12)$$

Note that $p_1$ and $p_l$ are nonnegative even numbers while $q_1$ and $q_l$ are positive odd numbers. Accordingly, $p_1 p_l$, $p_1 q_l$, and $q_1 p_l$ are nonnegative even numbers but $q_1 q_l$ are positive odd numbers. Thus, for $\chi_{1l} t = \pi$, we have $\exp(-ip_1 p_l \chi_{1l} t) = \exp(-ip_1 q_l \chi_{1l} t) = \exp(-iq_1 p_l \chi_{1l} t) = 1$ but $\exp(-iq_1 q_l \chi_{1l} t) = -1$. Hence, the state transformation (12) becomes

$$U_{1l}|0_1 0_l\rangle = |0_1 0_l\rangle, \quad U_{1l}|0_1 1_l\rangle = |0_1 1_l\rangle,$$
$$U_{1l}|1_1 0_l\rangle = |1_1 0_l\rangle, \quad U_{1l}|1_1 1_l\rangle = -|1_1 1_l\rangle. \quad (13)$$

This result (13) indicates that by applying a unitary operator $U_{1l}$, a universal controlled-phase gate on two photonic qubits 1 and $l$, described by $|0_1 0_l\rangle \to |0_1 0_l\rangle$, $|0_1 1_l\rangle \to |0_1 1_l\rangle$, $|1_1 0_l\rangle \to |1_1 0_l\rangle$, and $|1_1 1_l\rangle \to -|1_1 1_l\rangle$, is implemented. Note that the state transformation (13) can be simplified as

$$U_{1l}|0_1 i_l\rangle = |0_1 i_l\rangle, \quad U_{1l}|1_1 i_l\rangle = (-1)^{i_l}|1_1 i_l\rangle, \quad (14)$$

where $i_l \in \{0, 1\}$.

Based on Eq. (14), it is straightforward to get the following state transformation:

$$\prod_{l=2}^{n} U_{1l}|0_1\rangle|i_2\rangle|i_3\rangle \cdots |i_n\rangle = |0_1\rangle|i_2\rangle|i_3\rangle \cdots |i_n\rangle,$$

$$\prod_{l=2}^{n} U_{1l}|1_1\rangle|i_2\rangle|i_3\rangle \cdots |i_n\rangle = |1_1\rangle(-1)^{i_2}(-1)^{i_3} \cdots (-1)^{i_n}|i_2\rangle|i_3\rangle \cdots |i_n\rangle. \quad (15)$$

On the other hand, based on Eqs. (10) and (2), the unitary operator $U_1$, acting on the two logic states $|0_1\rangle$ and $|1_1\rangle$ of photonic qubit 1, results in the following state transformation:

$$U_1|0_1\rangle = \sum_{p_1} \exp(-ip_1\eta t)d_{p_1}|p_1\rangle,$$

$$U_1|1_1\rangle = \sum_{q_1} \exp(-iq_1\eta t)d_{q_1}|q_1\rangle. \quad (16)$$

For $\eta t = 2s\pi$ ($s$ is an integer), we have $\exp(-ip_1\eta t) = \exp(-iq_1\eta t) = 1$. Thus, it follows from Eq. (16) that

$$U_1|0_1\rangle = |0_1\rangle, \quad U_1|1_1\rangle = |1_1\rangle. \quad (17)$$

By combining Eq. (15) with Eq. (17), one can have the following state transformation:

$$U_1 \prod_{l=2}^{n} U_{1l}|0_1\rangle|i_2\rangle|i_3\rangle \cdots |i_n\rangle = |0_1\rangle|i_2\rangle|i_3\rangle \cdots |i_n\rangle,$$

$$U_1 \prod_{l=2}^{n} U_{1l}|1_1\rangle|i_2\rangle|i_3\rangle \cdots |i_n\rangle = |1_1\rangle(-1)^{i_2}(-1)^{i_3} \cdots (-1)^{i_n}|i_2\rangle|i_3\rangle \cdots |i_n\rangle. \quad (18)$$

Because of $U = U_1 \otimes \prod_{l=2}^{n} U_{1l}$ [see Eq. (9)], Eq. (18) can be simplified as

$$U|0_1\rangle|i_2\rangle|i_3\rangle \cdots |i_n\rangle = |0_1\rangle|i_2\rangle|i_3\rangle \cdots |i_n\rangle,$$

$$U|1_1\rangle|i_2\rangle|i_3\rangle \cdots |i_n\rangle = |1_1\rangle(-1)^{i_2}(-1)^{i_3} \cdots (-1)^{i_n}|i_2\rangle|i_3\rangle \cdots |i_n\rangle, \quad (19)$$

which implies that if the control photonic qubit 1 is in state $|0\rangle$, the states of each of target photonic qubit $(2, 3, ..., n)$ remain unchanged, while, if the control photonic qubit 1 is in state $|1\rangle$, a phase flip (from sign $+$ to $-$) happens to state $|1\rangle$ of each of target photonic qubit $(2, 3, ..., n)$. Hence, a multi-target-qubit controlled phase gate, described by Eq. (1), is realized with $n$ photonic qubits $(1, 2, ..., n)$, after the above operation.

### B. Discussion

From the description given above, one can see that:

(i) The unitary operator $U$ was obtained by starting the original Hamiltonian (3). Thus, the gate is implemented through a single-step operation which is described by $U$.

(ii) During the gate operation, the qutrit stays in the ground state. Thus, decoherence from the qutrit is greatly suppressed.

(iii) In the above, we have set $\chi_{1l}t = \pi$ (independent of $l$), $\eta t = 2s\pi$, and $\eta = -\lambda_1 + \sum_{l=2}^{n} \chi_{1l}$, which turn out to

$$-\lambda_1 t + (n-1)\pi = 2s\pi. \quad (20)$$

For even-number target qubits (i.e., $n - 1$ is even), one can choose $\lambda_1 t = 2m\pi$ ($m$ is a positive integer) in order to meet the condition given by Eq. (20). In this case, combining $\chi_{1l}t = \pi$ and $\lambda_1 t = 2m\pi$ results in $\chi_{1l} = \lambda_1/(2m)$, which turns into

$$g_l = \frac{\delta_l}{\delta_1 + \delta_l}\sqrt{2\Delta_{1l}\delta_1/m}. \quad (21)$$

On the other hand, for odd-number target qubits, one can select $\lambda_1 t = (2m' + 1)\pi$ ($m'$ is a nonnegative integer) to satisfy Eq. (20). In this case, combining $\chi_{1l}t = \pi$ and $\lambda_1 t = \pi$ results in $\chi_{1l} = \lambda_1/(2m' + 1)$, which leads to

$$g_l = \frac{2\delta_l}{\delta_1 + \delta_l}\sqrt{\Delta_{1l}\delta_1/(2m' + 1)}. \quad (22)$$

The condition (21) or (22) can be readily met by adjusting $g_l$ or $\delta_l$ or both, given $\delta_1$. It is noted that the detuning $\delta_l = \omega_{fe} - \omega_{c_l}$ can be tuned by changing the frequency $\omega_{c_l}$ of cavity $l$ ($l = 2, 3, ..., n$). In addition, the coupling strength $g_l$ can be adjusted by a prior design of the sample with appropriate capacitance or inductance between the coupler qutrit and cavity $l$ ($l = 2, 3, ..., n$) [95,96].

### C. Examples of photonic-qubit encodings

As stated previously, the two logic states of a photonic qubit are encoded with two arbitrary orthogonal eigenstates $|\varphi_e\rangle$ and $|\varphi_o\rangle$ of the photon-number parity operator $\hat{\pi} = e^{i\pi\hat{a}^+\hat{a}}$ of a cavity. Namely, to have the encoding be effective, the orthogonality between the two states $|\varphi_e\rangle$ and $|\varphi_o\rangle$ needs to hold, i.e.,

$$\langle \varphi_e | \varphi_o \rangle = 0. \quad (23)$$

In the following, we will provide a few examples of the encodings, for which the encoding condition (23) satisfies the following:

(i) The two logic states of each photonic qubit are encoded by the vacuum state $|0\rangle$ and the single-photon state $|1\rangle$, i.e., $|\varphi_e\rangle = |0\rangle$ and $|\varphi_o\rangle = |1\rangle$.

(ii) The two logic states of each photonic qubit are encoded by a Fock state $|2m\rangle$ with even-number photons and a Fock state $|2n + 1\rangle$ with odd-number photons, i.e., $|\varphi_e\rangle = |2m\rangle$ and $|\varphi_o\rangle = |2n + 1\rangle$ ($m$ and $n$ are nonnegative integers).

(iii) The two logic states of each photonic qubit are encoded by one superposition of Fock states with even-number photons and one superposition of Fock states with odd-number photons, i.e., $|\varphi_e\rangle = c_0|0\rangle + c_2|2\rangle + \cdots + c_{2m}|2m\rangle$ and $|\varphi_o\rangle = c_1|1\rangle + c_3|3\rangle + \cdots + c_{2n+1}|2n + 1\rangle$ ($m$ and $n$ are nonnegative integers).

(iv) The two logic states of each photonic qubit are encoded by two cat states, e.g., $|\varphi_e\rangle = |\alpha\rangle + |-\alpha\rangle$ and $|\varphi_o\rangle = |\alpha\rangle - |-\alpha\rangle$. Here, the cat state $|\alpha\rangle + |-\alpha\rangle$ is an even-number-photon coherent state while the cat state $|\alpha\rangle - |-\alpha\rangle$ is an odd-number-photon coherent state.

(v) The two logic states of each photonic qubit are encoded by one superposition of a Fock state and a cat state, each with even-number photons, and one superposition of a Fock state and a cat state, each with odd-number photons, e.g., $|\varphi_e\rangle = c_0|2m\rangle + c_1(|\alpha\rangle + |-\alpha\rangle)$ and $|\varphi_0\rangle = d_0|2n+1\rangle + d_1(|\alpha\rangle - |-\alpha\rangle)$ ($m$ and $n$ are nonnegative integers).

(vi) The two logic states of each photonic qubit are encoded by a squeezed vacuum state $|\xi\rangle$ and a cat state $|\alpha\rangle - |-\alpha\rangle$, i.e., $|\varphi_e\rangle = |\xi\rangle$ and $|\varphi_o\rangle = |\alpha\rangle - |-\alpha\rangle$. Note that the squeezed vacuum state is a superposition state of Fock states with even-number photons, while the cat state $|\alpha\rangle - |-\alpha\rangle$ is an odd-number-photon coherent state.

(vii) The two logic states of each photonic qubit are encoded by two multicomponent cat states, e.g., $|\varphi_e\rangle = c_0(|\alpha\rangle + |-\alpha\rangle) + c_1(|i\alpha\rangle + |-i\alpha\rangle)$ and $|\varphi_o\rangle = d_0(|\alpha\rangle - |-\alpha\rangle) + d_1(|i\alpha\rangle - |-i\alpha\rangle)$.

One can verify that for the examples above, the two states $|\varphi_e\rangle$ and $|\varphi_o\rangle$ are the eigenstates of the photon-number parity operator $e^{i\pi\hat{a}^+\hat{a}}$ (with eigenvalues $\pm 1$, respectively), and they satisfy the orthogonality condition (23). For simplicity, the states $|\varphi_e\rangle$ and $|\varphi_o\rangle$ given above for some examples are not normalized, which does not affect their orthogonality. The normalization of $|\varphi_e\rangle$ or $|\varphi_o\rangle$ can be easily made by adding a normalization coefficient. We should point out that, apart from the above examples, there are other possible encodings that satisfy the encoding condition (23). Therefore, the present proposal is quite general and can be used to implement the multi-target-qubit controlled-phase gate by using photonic qubits with various encodings, for which the two logic states used to encode photonic qubits can be any two orthogonal eigenstates of the photon-number parity operator.

### D. Nonhybrid and hybrid gates

The multi-target-qubit gate (19) or (1), whose implementation as shown in the preceding Sec. III A, can be a nonhybrid gate or a hybrid gate. In the case when the two logic states $|0\rangle$ and $|1\rangle$ of each photonic qubit are encoded by the same two arbitrary orthogonal eigenstates $|\varphi_e\rangle$ and $|\varphi_o\rangle$, i.e., $|\varphi_e\rangle_{c_1} = |\varphi_e\rangle_{c_2} = \cdots = |\varphi_e\rangle_{c_n}$ and $|\varphi_o\rangle_{c_1} = |\varphi_o\rangle_{c_2} = \cdots = |\varphi_o\rangle_{c_n}$, the gate (19) or (1) is a nonhybrid gate. Since the states $|\varphi_e\rangle$ and $|\varphi_o\rangle$ can take various quantum states, the present proposal can be used to implement *nonhybrid* multi-target-qubit gates using photonic qubits with various encodings.

On the other hand, if there exists at least one photonic qubit whose encoding is different from the encodings of other photonic qubits, the gate (19) or (1) is a hybrid gate. When the encodings of all photonic qubits are different, i.e., $|\varphi_e\rangle_{c_1} \neq |\varphi_e\rangle_{c_2} \neq \cdots \neq |\varphi_e\rangle_{c_n}$ and $|\varphi_o\rangle_{c_1} \neq |\varphi_o\rangle_{c_2} \neq \cdots \neq |\varphi_o\rangle_{c_n}$, we say that the multi-target-qubit gate (19) or (1) has a maximal hybridization. Because the states $|\varphi_e\rangle$ and $|\varphi_o\rangle$ can take various quantum states, the present proposal can also be applied to realize *hybrid* multi-target-qubit gates using photonic qubits with different encodings.

## IV. GENERATION OF MULTICAVITY GHZ ENTANGLED STATES

In this section, we discuss how to apply the gate (19) or (1) to generate a multi-cavity GHZ entangled state with general expression. Depending on the specific encodings $|\varphi_e\rangle$ and $|\varphi_o\rangle$, we then discuss the preparation of a few non-hybrid and hybrid GHZ entangled states of multiple cavities.

### A. Preparation of GHZ entangled state with general expression

Let us return to the physical system depicted in Fig. 2. Assume that the coupler SC qutrit is initially in the ground state $|g\rangle$ and each cavity is initially in the state $(|\varphi_e\rangle + |\varphi_o\rangle)/\sqrt{2}$. Here, the states $|\varphi_e\rangle$ and $|\varphi_o\rangle$ are normalized. The initial state of the whole system is thus given by

$$|\psi(0)\rangle = 2^{-n/2}\left(|\varphi_e\rangle_{c_1} + |\varphi_o\rangle_{c_1}\right)\prod_{l=2}^{n}\left(|\varphi_e\rangle_{c_l} + |\varphi_o\rangle_{c_l}\right) \otimes |g\rangle. \quad (24)$$

Here and after, the subscript $c_1$ represents cavity 1 while the subscript $c_l$ represents cavity $l$ ($l = 2, 3, ..., n$). In view of Eq. (2), i.e., based on the encoding $|0_k\rangle = |\varphi_e\rangle_{c_k}$ and $|1_k\rangle = |\varphi_o\rangle_{c_k}$, the state $|\psi(0)\rangle$ can be rewritten as

$$|\psi(0)\rangle = 2^{-n/2}(|0_1\rangle + |1_1\rangle)\prod_{l=2}^{n}(|0_l\rangle + |1_l\rangle) \otimes |g\rangle, \quad (25)$$

where states $|0_1\rangle$ and $|1_1\rangle$ are the two logic states of photonic qubit 1, while states $|0_l\rangle$ and $|1_l\rangle$ are the two logic states of photonic qubit $l$ ($l = 2, 3, ..., n$).

Now apply the multi-target-qubit controlled-phase gate (19), i.e., the gate described by Eq. (1). One can easily see that after this gate operation, the state (25) changes to

$$2^{-n/2}\left[|0_1\rangle\prod_{l=2}^{n}(|0_l\rangle + |1_l\rangle) + |1_1\rangle\prod_{l=2}^{n}(|0_l\rangle - |1_l\rangle)\right] \otimes |g\rangle. \quad (26)$$

According to the encoding (2), i.e., $|0_k\rangle = |\varphi_e\rangle_{c_k}$ and $|1_k\rangle = |\varphi_o\rangle_{c_k}$, the state (26) can be written as

$$2^{-n/2}\left[|\varphi_e\rangle_{c_1}\prod_{l=2}^{n}\left(|\varphi_e\rangle_{c_l} + |\varphi_o\rangle_{c_l}\right)\right.$$
$$\left. + |\varphi_o\rangle_{c_1}\prod_{l=2}^{n}\left(|\varphi_e\rangle_{c_l} - |\varphi_o\rangle_{c_l}\right)\right] \otimes |g\rangle, \quad (27)$$

which shows that the cavity system is prepared in the following GHZ entangled state with general expression

$$|\text{GHZ}\rangle = \frac{1}{\sqrt{2}}\left[|\varphi_e\rangle_{c_1}\prod_{l=2}^{n}\frac{1}{\sqrt{2}}\left(|\varphi_e\rangle_{c_l} + |\varphi_o\rangle_{c_l}\right)\right.$$
$$\left. + |\varphi_o\rangle_{c_1}\prod_{l=2}^{n}\frac{1}{\sqrt{2}}\left(|\varphi_e\rangle_{c_l} - |\varphi_o\rangle_{c_l}\right)\right]. \quad (28)$$

One can verify that for cavity $l$ ($l = 2, 3, ..., n$), the two rotated states $(|\varphi_e\rangle_l + |\varphi_o\rangle_l)/\sqrt{2}$ and $(|\varphi_e\rangle_l - |\varphi_o\rangle_l)/\sqrt{2}$ are orthogonal to each other.

From the descriptions given above, it can be seen that, independent of $|\varphi_e\rangle_l$ and $|\varphi_o\rangle_l$, the GHZ entangled state (28) of $n$ cavities $(1, 2, ..., n)$, which takes a general form, can be straightforwardly generated by applying the multi-target-qubit controlled-phase gate (19) or (1). Thus, given that the initial state $(|\varphi_e\rangle + |\varphi_o\rangle)/\sqrt{2}$ of each cavity is ready, the Hamiltonians and the operation used for preparing the GHZ state (28) are the same as those for implementing the gate (19) or (1). Moreover, as shown above, the gate is implemented with a single-step operation. Thus, the GHZ state (28) can be created through a single-step operation, given that the initial state of each cavity is ready.

## B. Preparation of nonhybrid and hybrid GHZ entangled states

In the following, we will show that, according to Eq. (28), which gives a general expression of the prepared GHZ entangled state of $n$ cavities, several nonhybrid and hybrid GHZ entangled states of $n$ cavities can be created by choosing appropriate encodings of $|\varphi_e\rangle_l$ and $|\varphi_o\rangle_l$.

To begin with, we define the following:

$$|+\rangle_{c_k} = (|0\rangle_{c_k} + |1\rangle_{c_k})/\sqrt{2},$$
$$|-\rangle_{c_k} = (|0\rangle_{c_k} - |1\rangle_{c_k})/\sqrt{2}, \quad (29)$$

where $|+\rangle_{c_k}$ and $|-\rangle_{c_k}$ are the two rotated states of cavity $k$ ($k = 1, 2, ..., n$). Note that $|0\rangle_{c_k}$ and $|1\rangle_{c_k}$ are the vacuum state and the single-photon state of cavity $k$, which are different from the two logic states $|0_k\rangle$ and $|1_k\rangle$ of photonic qubit $k$ ($k = 1, 2, ..., n$). In addition, we define

$$|cat\rangle_{c_k} = \mathcal{N}(|\alpha\rangle_{c_k} + |-\alpha\rangle_{c_k}),$$
$$|\overline{cat}\rangle_{c_k} = \mathcal{N}(|\alpha\rangle_{c_k} - |-\alpha\rangle_{c_k}), \quad (30)$$

where $|cat\rangle_{c_k}$ and $|\overline{cat}\rangle_{c_k}$ are the two cat states of cavity $k$ ($k = 1, 2, ..., n$), with a normalization coefficient $\mathcal{N} = 1/\sqrt{2(1 + e^{-2|\alpha|^2})}$.

### 1. Preparation of nonhybrid GHZ entangled state

Consider $|\varphi_e\rangle_{c_k} = |0\rangle_{c_k}$ and $|\varphi_o\rangle_{c_k} = |1\rangle_{c_k}$ ($k = 1, 2, ...n$). In this case, according to Eq. (29), it follows from Eq. (24) that the initial state of the system is

$$|\psi(0)\rangle = |+\rangle_{c_1}|+\rangle_{c_2} \cdots |+\rangle_{c_n} \otimes |g\rangle. \quad (31)$$

According to Eq. (28), the $n$ cavities are prepared in the following GHZ entangled state:

$$|\text{GHZ}\rangle = \frac{1}{\sqrt{2}}\left(|0\rangle_{c_1} \prod_{l=2}^{n} |+\rangle_{c_l} + |1\rangle_{c_1} \prod_{l=2}^{n} |-\rangle_{c_l}\right). \quad (32)$$

By performing a local operation on cavity $l$ and the coupler qutrit or applying a local operation on cavity $l$ and an auxiliary two-level SC qubit (placed in cavity $l$), one can achieve the state transformation $|+\rangle_{c_l} \to |0\rangle_{c_l}$ and $|-\rangle_{c_l} \to |1\rangle_{c_l}$ [16]. Thus, the GHZ state (32) turns into

$$|\text{GHZ}\rangle = \frac{1}{\sqrt{2}}(|0\rangle_{c_1}|0\rangle_{c_2} \cdots |0\rangle_{c_n} + |1\rangle_{c_1}|1\rangle_{c_2} \cdots |1\rangle_{c_n}), \quad (33)$$

which is a nonhybrid GHZ entangled state of $n$ cavities $(1, 2, ..., n)$.

### 2. Preparation of cat-coherent hybrid GHZ entangled state

Consider $|\varphi_e\rangle_{c_k} = |cat\rangle_{c_k}$ and $|\varphi_o\rangle_{c_k} = |\overline{cat}\rangle_{c_k}$ ($k = 1, 2, ..., n$). In this case, according to Eq. (30), it follows from Eq. (24) that the initial state of the system is

$$|\psi(0)\rangle = |\alpha\rangle_{c_1}|\alpha\rangle_{c_2} \cdots |\alpha\rangle_{c_n} \otimes |g\rangle, \quad (34)$$

i.e., each cavity is initially in a coherent state $|\alpha\rangle$. According to Eq. (28), the $n$ cavities are prepared in the following cat-coherent hybrid GHZ entangled state:

$$|\text{GHZ}\rangle = \frac{1}{\sqrt{2}}\left(|cat\rangle_{c_1} \prod_{l=2}^{n} |\alpha\rangle_{c_l} + |\overline{cat}\rangle_{c_1} \prod_{l=2}^{n} |-\alpha\rangle_{c_l}\right). \quad (35)$$

### 3. Preparation of cat-spin hybrid GHZ entangled state

For cavity 1, consider $|\varphi_e\rangle_{c_1} = |cat\rangle_{c_1}$ and $|\varphi_o\rangle_{c_1} = |\overline{cat}\rangle_{c_1}$. For cavity $k$ ($k = 2, 3, ..., n$), consider $|\varphi_e\rangle_{c_k} = |0\rangle_{c_k}$ and $|\varphi_o\rangle_{c_k} = |1\rangle_{c_k}$. In this case, according to Eqs. (29) and (30), it follows from Eq. (24) that the initial state of the system is

$$|\psi(0)\rangle = |\alpha\rangle_{c_1} \prod_{k=2}^{n} |+\rangle_{c_k} \otimes |g\rangle. \quad (36)$$

According to Eq. (28), the $n$ cavities are prepared in the following cat-spin hybrid GHZ entangled state:

$$|\text{GHZ}\rangle = \frac{1}{\sqrt{2}}\left(|cat\rangle_{c_1} \prod_{k=2}^{n} |+\rangle_{c_k} + |\overline{cat}\rangle_{c_1} \prod_{k=2}^{n} |-\rangle_{c_k}\right), \quad (37)$$

where the two rotated states $|+\rangle = (|0\rangle + |1\rangle)/\sqrt{2}$ and $|-\rangle = (|0\rangle - |1\rangle)/\sqrt{2}$ correspond to the spin left-hand circular motion and the spin right-hand circular motion, respectively.

### 4. Preparation of spin-coherent hybrid GHZ entangled state

Consider $|\varphi_e\rangle_{c_1} = |0\rangle_{c_1}$, $|\varphi_o\rangle_{c_1} = |1\rangle_{c_1}$, $|\varphi_e\rangle_{c_l} = |cat\rangle_{c_l}$, and $|\varphi_o\rangle_{c_l} = |\overline{cat}\rangle_{c_l}$ ($l = 2, 3, ..., n$). In this case, according to Eqs. (29) and (30), it follows from Eq. (24) that the initial state of the system is

$$|\psi(0)\rangle = |+\rangle_{c_1}|\alpha\rangle_{c_2}|\alpha\rangle_{c_3} \cdots |\alpha\rangle_{c_n} \otimes |g\rangle. \quad (38)$$

According to Eq. (28), the $n$ cavities are prepared in the following spin-coherent hybrid GHZ entangled state:

$$|\text{GHZ}\rangle = \frac{1}{\sqrt{2}}\left(|0\rangle_{c_1} \prod_{l=2}^{n} |\alpha\rangle_{c_l} + |1\rangle_{c_1} \prod_{l=2}^{n} |-\alpha\rangle_{c_l}\right). \quad (39)$$

Note that the vacuum state and the single-photon state of cavity 1 here correspond to the spin up and down, respectively.

To ensure that the prepared states (33), (35), (37), and (39) above are GHZ states, the two states $|cat\rangle$ and $|\overline{cat}\rangle$, $|+\rangle$ and $|-\rangle$, $|0\rangle$ and $|1\rangle$, or $|\alpha\rangle$ and $|-\alpha\rangle$ of each cavity are required to be orthogonal or quasi-orthogonal to each other. Note that the two cat states $|cat\rangle$ and $|\overline{cat}\rangle$ are orthogonal, the two rotated states $|+\rangle$ and $|-\rangle$ are orthogonal, and the two states $|0\rangle$ and $|1\rangle$ are orthogonal. In addition, the two coherent states $|\alpha\rangle$ and $|-\alpha\rangle$ can be made to be quasi-orthogonal for a large enough $\alpha$ (e.g., $|\langle\alpha|-\alpha\rangle|^2 = e^{-4|\alpha|^2} \lesssim 10^{-2}$ for $\alpha = 1.1$).

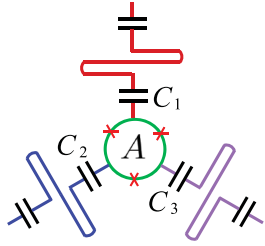

FIG. 4. Diagram of three 1D microwave cavities capacitively coupled to a SC flux qutrit. Each cavity here is a one-dimensional transmission line resonator. The flux qutrit consists of three Josephson junctions and a superconducting loop.

Before ending this section, we should mention that, in addition to the above GHZ states (33), (35), (37), and (39), other types of nonhybrid or hybrid GHZ entangled states of multiple cavities can be created according to Eq. (28), because there exist other possible encodings of $|\varphi_e\rangle$ and $|\varphi_o\rangle$. The multi-target-qubit controlled-phase gate considered in this work can be used to prepare various nonhybrid and hybrid GHZ entangled states of multiple cavities.

## V. POSSIBLE EXPERIMENTAL IMPLEMENTATION

In this section, as an example, we investigate the experimental feasibility for creating a hybrid GHZ state of three cavities (1,2,3), by using three 1D microwave cavities coupled to a SC flux qutrit (Fig. 4). The hybrid GHZ state to be prepared is

$$|\text{GHZ}\rangle = \frac{1}{\sqrt{2}}\big(|0\rangle_{c_1}|\alpha\rangle_{c_2}|\alpha\rangle_{c_3} + |1\rangle_{c_1}|-\alpha\rangle_{c_2}|-\alpha\rangle_{c_3}\big), \quad (40)$$

i.e., a spin-coherent hybrid GHZ state, given by Eq. (39) for $n=3$. From Eq. (38), one can see that the initial state of the whole system is

$$|\psi(0)\rangle = |+\rangle_{c_1}|\alpha\rangle_{c_2}|\alpha\rangle_{c_3} \otimes |g\rangle. \quad (41)$$

### A. Full Hamiltonian

As discussed in Sec. IV A, one can see that the GHZ state (40) here is prepared by applying the gate (19) or (1) with $n=3$ (i.e., a three-qubit gate with one qubit simultaneously controlling two target qubits). Thus, the Hamiltonian used for the generation of the GHZ state (40) is the effective Hamiltonian (7) with $n=3$. This Hamiltonian was derived from the original Hamiltonian (3) with $n=3$, which only includes the coupling between cavity 1 and the $|g\rangle \leftrightarrow |f\rangle$ transition as well as the coupling between other cavities and the $|e\rangle \leftrightarrow |f\rangle$ transition of the SC qutrit. In a realistic situation, there is the unwanted coupling between cavity 1 and the $|e\rangle \leftrightarrow |f\rangle$ transition, the unwanted coupling between cavity 2 and the $|g\rangle \leftrightarrow |f\rangle$ transition, as well as the unwanted coupling between cavity 3 and the $|g\rangle \leftrightarrow |f\rangle$ transition of the SC qutrit. Besides, there is the unwanted intercavity crosstalk between the three cavities.

When the unwanted couplings and the unwanted intercavity crosstalk are taken into account, the Hamiltonian (3), with

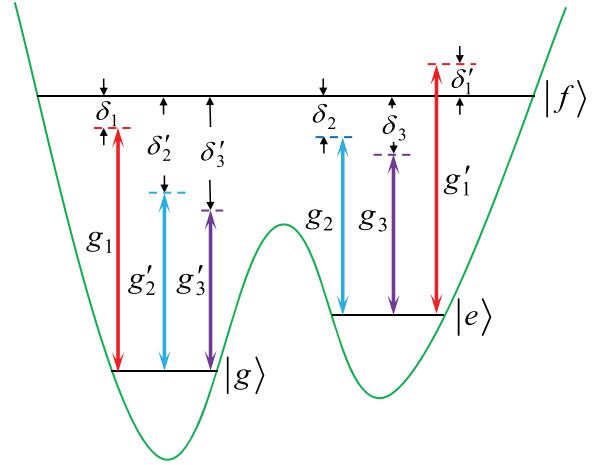

FIG. 5. Illustration of the required coupling between cavity 1 and the $|g\rangle \leftrightarrow |f\rangle$ transition of the qutrit (with coupling constant $g_1$ and detuning $\delta_1$), the required coupling between cavity 2 and the $|e\rangle \leftrightarrow |f\rangle$ transition of the qutrit (with coupling constant $g_2$ and detuning $\delta_2$), and the required coupling between cavity 3 and the $|e\rangle \leftrightarrow |f\rangle$ transition of the qutrit (with coupling constant $g_3$ and detuning $\delta_3$). In addition, illustration of the unwanted coupling between cavity 2 and the $|g\rangle \leftrightarrow |f\rangle$ transition of the qutrit (with coupling constant $g'_2$ and detuning $\delta'_2$), the unwanted coupling between cavity 1 and the $|e\rangle \leftrightarrow |f\rangle$ transition of the qutrit (with coupling constant $g'_1$ and detuning $\delta'_1$), the unwanted coupling between cavity 2 and the $|g\rangle \leftrightarrow |f\rangle$ transition of the qutrit (with coupling constant $g'_2$ and detuning $\delta'_2$), as well as the unwanted coupling between cavity 3 and the $|g\rangle \leftrightarrow |f\rangle$ transition of the qutrit (with coupling constant $g'_3$ and detuning $\delta'_3$). Note that the coupling of each cavity with the $|g\rangle \leftrightarrow |e\rangle$ transition of the qutrit is negligible because of the weak $|g\rangle \leftrightarrow |e\rangle$ transition. Red, blue, and purple lines correspond to cavities 1, 2, and 3, respectively.

$n=3$ for the present case, is modified as

$$\begin{aligned}H'_I =\ & g_1(e^{-i\delta_1 t}\hat{a}_1^+ \sigma_{fg}^- + \text{H.c.}) + \sum_{l=2}^{3} g_l(e^{-i\delta_l t}\hat{a}_l^+ \sigma_{fe}^- + \text{H.c.})\\& + g'_1(e^{-i\delta'_1 t}\hat{a}_1^+ \sigma_{fe}^- + \text{H.c.}) + \sum_{l=2}^{3} g'_l(e^{-i\delta'_l t}\hat{a}_l^+ \sigma_{fg}^- + \text{H.c.}),\\& + (\widetilde{g}_{12} e^{i\widetilde{\Delta}_{12} t}\hat{a}_1^+\hat{a}_2 + \widetilde{g}_{13} e^{i\widetilde{\Delta}_{23} t}\hat{a}_2^+\hat{a}_3\\& + \widetilde{g}_{23} e^{i\widetilde{\Delta}_{13} t}\hat{a}_1^+\hat{a}_3 + \text{H.c.}),\end{aligned} \quad (42)$$

where the terms in line one represent the required coupling between cavity 1 and the $|g\rangle \leftrightarrow |f\rangle$ transition as well as the required coupling between cavity 2 (cavity 3) and the $|e\rangle \leftrightarrow |f\rangle$ transition of the SC qutrit (Fig. 5), the first term in line two represents the unwanted coupling between cavity 1 and the $|e\rangle \leftrightarrow |f\rangle$ transition of the SC qutrit with coupling constant $g'_1$ and detuning $\delta'_1 = \omega_{fe} - \omega_{c_1}$ (Fig. 5), the second term in line two represents the unwanted coupling between cavity 2 (cavity 3) and the $|g\rangle \leftrightarrow |f\rangle$ transition of the SC qutrit with coupling constant $g'_2$ ($g'_3$) and detuning $\delta'_2 = \omega_{fg} - \omega_{c_2}$ ($\delta'_3 = \omega_{fg} - \omega_{c_3}$) (Fig. 5), while the terms in the last line represent the unwanted intercavity crosstalk among the three cavities, with $\widetilde{g}_{kl}$ ($\widetilde{\Delta}_{kl} = \omega_{c_k} - \omega_{c_l}$) the crosstalk strength (the

TABLE I. Parameters used in the numerical simulation. $\omega_{eg}$, $\omega_{fe}$, and $\omega_{fg}$ are the $|g\rangle \leftrightarrow |e\rangle$, $|e\rangle \leftrightarrow |f\rangle$, and $|g\rangle \leftrightarrow |f\rangle$ transition frequencies of the flux qutrit, respectively. $\omega_{c_j}$ is the frequency of cavity $j$ ($j = 1, 2, 3$). $\delta_1$ ($\delta_1'$) is the detuning between the frequency of cavity 1 and the $|g\rangle \leftrightarrow |f\rangle$ ($|e\rangle \leftrightarrow |f\rangle$) transition frequency of the flux qutrit. $\delta_2$ ($\delta_2'$) is the detuning between the frequency of cavity 2 and the $|e\rangle \leftrightarrow |f\rangle$ ($|g\rangle \leftrightarrow |f\rangle$) transition frequency of the flux qutrit. $\delta_3$ ($\delta_3'$) is the detuning between the frequency of cavity 3 and the $|e\rangle \leftrightarrow |f\rangle$ ($|g\rangle \leftrightarrow |f\rangle$) transition frequency of the flux qutrit. $\widetilde{\Delta}_{kl}$ is the frequency detuning between the two cavities $k$ and $l$ ($kl = 12, 13, 23$). $g_1$ ($g_1'$) is the coupling constant between cavity 1 and the $|g\rangle \leftrightarrow |f\rangle$ ($|e\rangle \leftrightarrow |f\rangle$) transition of the flux qutrit. $g_2$ ($g_2'$) is the coupling constant between cavity 2 and the $|e\rangle \leftrightarrow |f\rangle$ ($|g\rangle \leftrightarrow |f\rangle$) transition of the flux qutrit. $g_3$ ($g_3'$) is the detuning between cavity 3 and the $|e\rangle \leftrightarrow |f\rangle$ ($|g\rangle \leftrightarrow |f\rangle$) transition of the flux qutrit.

| | | |
|---|---|---|
| $\omega_{eg}/(2\pi) = 8.0\,\text{GHz}$ | $\omega_{fe}/(2\pi) = 12.0\,\text{GHz}$ | $\omega_{fg}/(2\pi) = 20.0\,\text{GHz}$ |
| $\omega_{C1}/(2\pi) = 18.6\,\text{GHz}$ | $\omega_{C2}/(2\pi) = 10.0\,\text{GHz}$ | $\omega_{C3}/(2\pi) = 9.6\,\text{GHz}$ |
| $\delta_1/(2\pi) = 1.6\,\text{GHz}$ | $\delta_2/(2\pi) = 2.0\,\text{GHz}$ | $\delta_3/(2\pi) = 2.4\,\text{GHz}$ |
| $\delta_1'/(2\pi) = -6.6\,\text{GHz}$ | $\delta_2'/(2\pi) = 10.0\,\text{GHz}$ | $\delta_3'/(2\pi) = 10.4\,\text{GHz}$ |
| $\widetilde{\Delta}_{12}/(2\pi) = 8.6\,\text{GHz}$ | $\widetilde{\Delta}_{13}/(2\pi) = 0.4\,\text{GHz}$ | $\widetilde{\Delta}_{23}/(2\pi) = 9.0\,\text{GHz}$ |
| $g_1/(2\pi) = 0.16\,\text{GHz}$ | $g_2/(2\pi) = 0.198\,\text{GHz}$ | $g_3/(2\pi) = 0.303\,\text{GHz}$ |
| $g_1'/(2\pi) = 0.16\,\text{GHz}$ | $g_2'/(2\pi) = 0.198\,\text{GHz}$ | $g_3'/(2\pi) = 0.303\,\text{GHz}$ |

frequency detuning) between the two cavities $k$ and $l$ ($k, l \in \{1, 2, 3\}; k \neq l$). Note that the coupling of each cavity with the $|g\rangle \leftrightarrow |e\rangle$ transition of the SC qutrit is negligible because the $|g\rangle \leftrightarrow |e\rangle$ transition can be made weak by increasing the barrier between the two potential wells of the SC qutrit. Hence, this coupling is not considered in the above Hamiltonian (42) to simplify our numerical simulations.

### B. Numerical results

With finite qutrit relaxation, dephasing and photon lifetime being included, the dynamics of the lossy system is determined by the following master equation:

$$\begin{aligned}
\frac{d\rho}{dt} = &-i[H_I', \rho] + \sum_{j=1}^{3} \kappa_j \mathcal{L}[\hat{a}_j] \\
&+ \gamma_{eg}\mathcal{L}[\sigma_{eg}^-] + \gamma_{fe}\mathcal{L}[\sigma_{fe}^-] + \gamma_{fg}\mathcal{L}[\sigma_{fg}^-] \\
&+ \gamma_{e,\varphi}(\sigma_{ee}\rho\sigma_{ee} - \sigma_{ee}\rho/2 - \rho\sigma_{ee}/2) \\
&+ \gamma_{f,\varphi}(\sigma_{ff}\rho\sigma_{ff} - \sigma_{ff}\rho/2 - \rho\sigma_{ff}/2),
\end{aligned} \quad (43)$$

where $\sigma_{ee} = |e\rangle\langle e|$, $\sigma_{ff} = |f\rangle\langle f|$, $\mathcal{L}[\Lambda] = \Lambda\rho\Lambda^+ - \Lambda^+\Lambda\rho/2 - \rho\Lambda^+\Lambda/2$ (with $\Lambda = \hat{a}_j, \sigma_{eg}^-, \sigma_{fe}^-, \sigma_{fg}^-$), $\gamma_{eg}$ is the energy relaxation rate of the level $|e\rangle$ for the decay path $|e\rangle \rightarrow |g\rangle$, $\gamma_{fe}$ ($\gamma_{fg}$) is the relaxation rate of the level $|f\rangle$ for the decay path $|f\rangle \rightarrow |e\rangle$ ($|f\rangle \rightarrow |g\rangle$), $\gamma_{e,\varphi}$ ($\gamma_{f,\varphi}$) is the dephasing rate of the level $|e\rangle$ ($|f\rangle$) of the qutrit, while $\kappa_j$ is the decay rate of cavity $j$ ($j = 1, 2, 3$). For numerical calculations, we here use the QUTIP software [97,98], which is an open-source software for simulating the dynamics of open quantum systems.

The fidelity for the prepared three-cavity GHZ state is evaluated by

$$\mathcal{F} = \sqrt{\langle\psi_{id}|\rho|\psi_{id}\rangle}, \quad (44)$$

where $|\psi_{id}\rangle = |\text{GHZ}\rangle \otimes |g\rangle$ is the ideal output state, with the state $|\text{GHZ}\rangle$ given in Eq. (40). The ideal output state $|\psi_{id}\rangle$ here is obtained without taking into account the system dissipation, the intercavity crosstalk, and the unwanted couplings, while $\rho$ is the density operator describing the real output state of the system, which is achieved by numerically solving the master equation and considering the operation performed in a realistic situation.

For a flux qutrit, the typical transition frequency between adjacent energy levels can be made as 1–20 GHz [99–101]. As an example, consider the parameters listed in Table I, which are used in the numerical simulations. The coupling constants $g_2$ and $g_3$ in Table I are calculated for $m = 10$ according to Eq. (21). By a proper design of the flux qutrit [102], one can have $\phi_{fg} \sim \phi_{fe} \sim 10\phi_{eg}$, where $\phi_{ij}$ is the dipole coupling matrix element between the two levels $|i\rangle$ and $|j\rangle$ with $ij \in \{eg, fe, fg\}$. Thus, one has $g_1' \sim g_1$, $g_2' \sim g_2$, and $g_3' \sim g_3$, while the $|g\rangle \leftrightarrow |e\rangle$ transition is much weaker compared to the $|g\rangle \leftrightarrow |f\rangle$ and $|e\rangle \leftrightarrow |f\rangle$ transitions of the qutrit. The maximum among the coupling constants listed in Table I is $g_{\max} = 2\pi \times 0.303$ GHz, which is readily available since a coupling constant $\sim 2\pi \times 0.636$ GHz has been reported for a flux device coupled to a microwave cavity [103].

Other parameters used in the numerical simulations are as follows: (i) $\gamma_{eg}^{-1} = 10T$, $\gamma_{fe}^{-1} = T$, $\gamma_{fg}^{-1} = T$, $\gamma_{\phi e}^{-1} = \gamma_{\phi f}^{-1} = T/2$, (ii) $\kappa_1 = \kappa_2 = \kappa_3 = \kappa$, (iii) $g_{12} = g_{23} = g_{13} = g_{cr}$, and (iv) $\alpha = 1.1$. We should mention that $\gamma_{eg}^{-1}$ is much larger than $\gamma_{fe}^{-1}$ and $\gamma_{fg}^{-1}$ because of $\phi_{fg} \sim \phi_{fe} \sim 10\phi_{eg}$. The maximal value of $T$ adopted in our numerical simulations is 20 μs. As a result, the decoherence times of the qutrit employed in the numerical simulations are 10−200 μs, which is a rather conservative case since a decoherence time of 70 μs to 1 ms for a superconducting flux device has been experimentally demonstrated [48,104].

By numerically solving the master equation (43), we plot Fig. 6, which gives the fidelity versus $\kappa^{-1}$ for $T = 10, 15,$ and 20 μs and $g_{cr} = 0.01g_{\max}$. The setting $g_{cr} = 0.01g_{\max}$ here is obtainable in experiments by a prior design of the sample with appropriate capacitances $C_1, C_2,$ and $C_3$ depicted in Fig. 4 [95]. One can see from Fig. 6 that the fidelity exceeds 93.07% for $\kappa^{-1} \geqslant 20$ μs and $T \geqslant 10$ μs. In addition, Fig. 6 shows that the fidelity is insensitive to the qutrit decoherence. This is because during the GHZ state preparation the qutrit mostly stays in the ground state and thus the effect of decoherence from the qutrit is negligible, while Fig. 6 shows that the fidelity is sensitive to the cavity decay. This is understood because the photons are populated in each cavity during the GHZ state preparation.

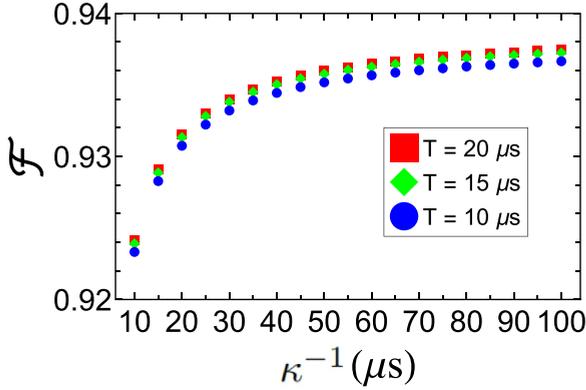

FIG. 6. Fidelity versus $\kappa^{-1}$ for $T = 10$, $15$, and $20\,\mu$s.

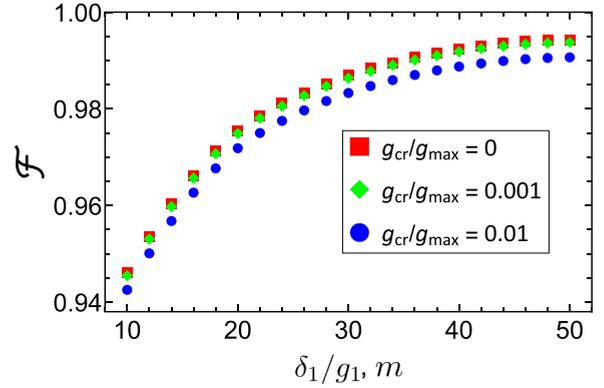

FIG. 8. Fidelity versus $\delta_1/g_1$ and $m$. Here, $m$ is the parameter involved in Eq. (21). The figure is plotted by setting $\delta_1/g_1 = m$, $g_1/2\pi = 0.16$ GHz, $\delta_2 = \delta_1 + 2\pi \times 0.4$ GHz, $\delta_3 = \delta_1 + 2\pi \times 0.8$ GHz, and assuming that the dissipation of the system, the unwanted couplings of each cavity with the transitions between the irrelevant levels of the qutrit (see Fig. 6), as well as the error in the initial state preparation are negligible. The coupling constants $g_2$ and $g_3$ used in the numerical simulation are calculated based on Eq. (21).

In reality, cavity 1 may not be prepared in a perfect initial state. Thus, we consider a nonideal initial state of the system

$$|\psi(0)\rangle_{\text{non-ideal}} = \mathcal{N}_x[(1+x)|0\rangle_{c_1} + (1-x)]|1\rangle_{c_1}|\alpha\rangle_{c_2}|\alpha\rangle_{c_3} \otimes |g\rangle, \quad (45)$$

where $\mathcal{N}_x = 1/\sqrt{2(1+x^2)}$ is a normalization factor. In this situation, we numerically plot Fig. 7 for $T = 10\,\mu$s, which shows that the fidelity decreases with increasing of $x$. Nevertheless, for $x \in [-0.1, 0.1]$, i.e., a 10% error in the weights of the $|0\rangle$ and $|1\rangle$ states, a fidelity greater than 92.48%, 92.92%, and 93.07% can be obtained for $\kappa^{-1} = 20$, $50$, and $100\,\mu$s, respectively.

With the parameters listed in Table I, the operational time for preparing the GHZ state is estimated to be $\sim 0.63\,\mu$s, much shorter than the cavity decay time and the qutrit decoherence time used in the numerical simulations. For the cavity frequencies given in Table I and $\kappa^{-1} = 20\,\mu$s, the quality factors for the three cavities are $Q_1 \sim 2.34 \times 10^6$, $Q_2 \sim 1.26 \times 10^6$, and $Q_3 \sim 1.21 \times 10^6$, which are achievable because a 1D microwave cavity or resonator with a high-quality factor $Q \gtrsim 2.7 \times 10^6$ has been reported in experiments [52,53].

### C. Discussion

The above analysis shows that the fidelity is sensitive to the error in the initial state and the cavity decay, but it is insensitive to the decoherence from the qutrit. The imperfect fidelity is also induced by the unwanted qutrit-cavity couplings. Moreover, the imperfect fidelity is caused because the large detuning condition is not well satisfied. We note that the fidelity can be further improved by optimizing the system parameters to reduce the errors. As demonstrated in Fig. 8, a high fidelity greater than 99% can be achieved for $\delta_1/g_1, m \gtrsim 50$, and $g_{cr} = 0.01 g_{\text{max}}$ when the unwanted qutrit-cavity couplings, the dissipation of the system, and the error in the initial state are negligible. Finally, it should be noted that further study is needed for each particular experimental setup. However, this requires a rather lengthy and complex analysis, which is beyond the scope of this theoretical work.

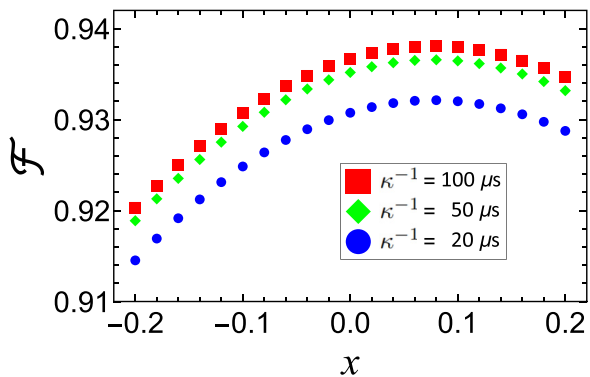

FIG. 7. Fidelity versus $x$ for $\kappa^{-1} = 20$, $50$, and $100\,\mu$s. In the numerical simulation, we set $T = 10\,\mu$s.

## VI. CONCLUSIONS

We have proposed a single-step method to implement a multi-target-qubit controlled-phase gate with photonic qubits each encoded via two arbitrary orthogonal eigenstates of the photon-number parity operator. The gate is realized by employing multiple microwave cavities coupled to a SC flux qutrit. As shown above, this proposal has the following features and advantages: (i) It can be applied not only to implement nonhybrid multi-target-qubit controlled-phase gates using photonic qubits with various encodings, but also to realize hybrid multi-target-qubit controlled-phase gates using photonic qubits with different encodings; (ii) the gate realization is quite simple because only a single-step operation is needed; (iii) neither classical pulse nor measurement is required, thus the gate is implemented in a deterministic way; (iv) since only one coupler SC qutrit is needed to couple all of the cavities, the hardware circuit resources are significantly reduced and minimized; (v) during the gate realization, the SC qutrit remains in the ground state, thus decoherence from the qutrit is greatly suppressed; and (vi) the operation time for the gate realization is independent of the number of target qubits, therefore it does not increase with increasing of the number of target qubits.

As an application, we have further discussed how to apply this multiqubit gate to generate a multicavity GHZ entangled state with general expression. Depending on the specific encodings, we have also discussed the generation of several nonhybrid and hybrid GHZ entangled states of multiple cavities. Furthermore, we have numerically analyzed the experimental feasibility of generating a spin-coherent hybrid GHZ state of three cavities within current circuit QED technology. To the best of our knowledge, this work is the first to demonstrate the realization of the proposed multiqubit gate with photonic qubits, encoded via two arbitrary orthogonal eigenstates of the photon-number parity operator, based on cavity or circuit QED. Finally, note that the same Hamiltonians introduced in the above Sec. III A can be obtained for many physical systems, such as multiple microwave or optical cavities coupled to a three-level natural or artificial atom. Hence, this proposal is quite general, thus can be applied to implement the proposed multi-target-qubit gate using photonic qubits with various encodings and can be used to prepare different kinds of nonhybrid and hybrid GHZ entangled states of multiple cavities in a wide range of physical systems.

### ACKNOWLEDGMENTS

This work was partly supported by the National Natural Science Foundation of China (NSFC) (Grants No. 11074062, No. 11374083, No. 11774076, and No. U21A20436), the Key-Area Research and Development Program of GuangDong province (Grant No. 2018B030326001), the Jiangsu Funding Program for Excellent Postdoctoral Talent, and the Innovation Program for Quantum Science and Technology (Grant No. 2021ZD0301705).